\documentstyle[12pt]{article}
\newtheorem{defi}{Definition}

\newtheorem{thm}[defi]{Theorem}

\makeatletter
\newenvironment{pf}%
 {\par\noindent{ Proof \quad}}{\hfill$\Box$\bigskip}
\makeatother

\def\tr{\mathop{\rm tr}\nolimits}
\def\Tr{\mathop{\rm Tr}\nolimits}

\def\vliminf{\mathop{\underline{\rm lim}}}

\def\argmax{\mathop{\rm argmax}}

\def\R{{\bf R}}

\def\Label#1{\label{#1}\ [\ #1\ ]\ }
\let\Label=\label

\begin{document}
\bibliographystyle{unsrt}
\vskip 2em \begin{center}
 {\LARGE\bf 
    Statistical Model with Measurement Degree of Freedom  
      and Quantum Physics
\par}
 \vskip 1.5em 
 \lineskip .5em
{  $\mbox{Masahito Hayashi}^{1}$ {\tt  masahito@qci.jst.go.jp}
\hspace{4ex}
\par
 $\mbox{Keiji Matsumoto}^{1, 2}$ {\tt keiji@nii.ac.jp}
\hspace{4ex}
 }\par
\end{center}

\footnotetext[1]{
Quantum Computation and Information Project, ERATO, JST,
        5-28-3 Hongo, Bunkyo-ku, Tokyo 113-0033, Japan.}
\footnotetext[2]{
National Institute of Informatiocs,
        Chiyoda-ku, Tokyo, Japan.}

\abstract{ 
This is an English translation of the manuscript[1] which was appeared in Surikaiseki Kenkyusho Kokyuroku No. 1055 (1998).
The asymptotic efficiency of statistical estimate of 
unknown quantum states is discussed, both in adaptive an
collective settings.  Aaptive bounds 
are written in sigle letterized form,
and collective bounds are written in limitting expression.
Our arguments clarify mathematical regularity conditions.}

\section{Introduction}
This is an English translation of the manuscript\cite{original}
 which appeared in Surikaiseki Kenkyusho Kokyuroku No. 1055 (1998).

In the estimation of
the unknown density operator 
by use of the experimental data,
the error can be reduced by 
the improvement of the desingn of the experiment.
Therefore, it is natural to ask what is the limit of the improvement.
To answer the question, Helstrom\cite{Helstrom:1976} founded
the quantum estimation theory,
in analogy with classical estimation theory(in the manuscript, we
reffer to  statistical estimation theory of probability distribution 
as `classical estimation theory').
Often, for simplisity, it is assumed that 
a state  belongs to a family 
${\cal M}
=\{\rho_\theta|\theta\in \Theta\subset {\R}^m\}$ 
of states,
which is called {\it model} and that 
the finite dimensional parameter $\theta$ 
is to be estimated statistically.

He considered the quantum analogue of 
 Cram{\'e}r-Rao inequality, which 
gives the lower bound of 
mean square error of locally unbiased estimate. 
This bound, however, is not achievable at all,
when the number of the data is finite.

Let us assume that the number $n$ of the data tends to
infinite. Then, if some regularity conditions are assumed,
it is concluded that if the estimate is consistent,
{\it i.e.},  the estimate converges to the true value of
parameter, the first order asymptotic term of
mean square error satisfies the Cram{\'e}r-Rao inequality,
and that the bound is achieved for all $\theta\in\Theta$.
This kind of discussion is called {\it first order asymptotic theory}.

The quatntum version of first order asymptotic theory is 
started by H. Nagaoka\cite{Nagaoka:1989:2}\cite{Nagb}. 
He defined, in our terminology,
the quasi-quantum Cram{\'e}r-Rao type bound,
and pointed out that the bound is achieved asymptotically and globally.
The proof of achivability, however, is only roughly sketched 
in his paper. 
In this manuscript, the proof of the achievability of the bound is 
fully written out, and the regularity conditions fo the achievability
is revealed. In addition, we defined another bound,
the quantum Cram{\'e}r-Rao type bound, and showed that the new bound is
also achievable, if the use of quantum correlation between samples are
allowed.

\section{Preliminaries}
An estimate $\hat\theta$ is obtained 
as a function $\hat\theta(\omega)$  of data $\omega\in \Omega$ to ${\rm R}^m$.
The purpose of the theory is to 
obtain the best estimate and its accuracy.
The optimization is done by the appropriate choice of
the measuring apparatus and the function $\hat\theta(\omega)$ from
data to the estimate.

Let $\sigma({\R}^m)$ be a $\sigma$- field in the space ${\R}^m$.
Whatever apparatus is used,
the data $\omega\in\Omega$
lie in a measurable subset $B\in \sigma({\R}^m)$
of $\Omega$ writes
\begin{eqnarray}
{\rm Pr}\{ \omega \in B|\theta \} =\tr \rho( \theta )M(B),
\Label{eqn:pdm}
\end{eqnarray}
when the true value of the parameter is $\theta$.
Here, $M$, which is called {\it positive operator valued measure
(, POM, in short)},
is a mapping  
from subsets $B\subset \Omega$ to non-negative Hermitian operators in 
${\cal H}$, such that
\begin{eqnarray}
&&M(\phi)=O, M(\Omega)=I,\nonumber\\
&&M(\bigcup_{i=1}^{\infty} B_i),
=\sum_{i=1}^{\infty}M(B_i)
\;\;(B_i\cap B_j=\phi,i\neq j),
\Label{eqn:maxiom}
\end{eqnarray}
(see Ref.{\rm \cite{Helstrom:1976}},p.53 
and Ref.{\rm \cite{Holevo:1982}},p.50.).
Conversely, some apparatus corresponds to any POM $M$
\cite{Steinspring:1955}\cite{Ozawa:1984}.
Therefore, we refer to the measurement which is controled by
the POM $M$ as `measurement $M$'.
A pair $(\hat\theta,M)$
is called an {\it estimator}. 

The classical Fisher information matrix $J^M_\theta$ by the POM $M$
is defined, as in the classical estimation theory,
\begin{eqnarray*}
 J^M_\theta:=
 \left[\int_{\omega\in\Omega} 
   \partial_i\log\frac{{\rm d}{\rm P}^M_\theta}{{\rm d}\nu}
     \partial_j\log\frac{{\rm d}{\rm P}^M_\theta}{{\rm d}\nu}
   {\rm dP}\right],
\end{eqnarray*}
where $\partial_i=\partial/\partial\theta^i$, 
${\rm P}^M_\theta(B):=\tr\rho_\theta M(B)$, and $\nu$ is 
some underlying measure 
(in the manuscript, we assume that for any POM $M$, 
 there is a measure $\nu$ in $\Omega$ such that
 ${\rm P}^M_\theta\prec \nu$ for all $\theta\in\Theta$).
Denote the mean square error matrix of $(\hat\theta M)$ by 
${\rm V}_{\theta}[\hat\theta, M]$,
and, as the measure of accuracy, let us take ${\rm Tr}G {\rm V}_{\theta}[M]$,
where $G$ is nonnegative symmetric real matrix.
If $G={\rm diag}(g_1,\cdots.g_m)$,  
${\rm Tr}G {\rm V}_{\theta}[M]$ is weighed sum of mean square error 
of the estimate $\hat\theta^i$ of each component $\theta^i$
of the parameter.

Let us define locally unbiased estimator $(\hat\theta,\,M)$ at $\theta$ by,
\begin{eqnarray}
 {\rm E}_{\theta}[\hat\theta_n,\,M]:=
\int\hat\theta^j(\omega) \tr\rho_\theta\, M({\rm d}\omega)
&=&\theta^j,\quad(j=1,\cdots,m).\\
 \int\hat\theta^j(\omega) \tr\partial_k\rho_\theta\, 
  M({\rm d}\omega)
   &=&\delta^j_k,\quad(j,k=1,\cdots,m).
\label{eqn:local_unbiasedness}
\end{eqnarray}
Then, $J^M_\theta$ is caracterized by,
\begin{eqnarray*}
 J^{M-1}_\theta=\inf\{V_\theta[\hat\theta,\,M]\,|\,
\mbox{$\hat\theta:(\hat\theta,\,M)$ is locally unbiased}\},
\end{eqnarray*}
and the quasi-quantum{\it Cram{\'e}r-Rao type bound} 

$C_\theta(G)$ is defined by,
\begin{eqnarray*}
C_\theta(G)&:=&\inf\{{\rm Tr}G{\rm V}_{\theta}[\hat\theta,\,M]\:|\:
                   \mbox{$M$ is locally unbiased}\}\nonumber\\
            &=&\inf\{{\rm Tr}G J^{M-1}_\theta\:|\: 
                          \mbox{$M$ is a POM in ${\cal H}$}\}.
\end{eqnarray*}
Nagaoka pointed out that 
the quasi-quantum{\it Cram{\'e}r-Rao type bound} is achievable
asymtotically 
for every $\theta\in\Theta$\cite{Nagaoka:1989:2}.
$C_\theta(G)$ is calculated explicitely for several special cases
\cite{Matu}\cite{Haya}.

Suppose $n$-i.i.d. pairs $\rho_\theta^{\otimes n}$ of the unknown state
$\rho_\theta$ are given. 
The  the sequence  $\{(\hat\theta,\,M_n)\}$, where
$M_n$ is a POM  in ${\cal H}^{\otimes n}$,
is said to be {\it MSE consistent} if the estimate $\hat\theta_n$ by
 converges to the true value of the parameter
in the mean, {i.e.},
$\lim_{n\to\infty} V_\theta[(\hat\theta_n,\,M_n)]=0$.

\section{The quasi-classical Cram{\'e}r-Rao type bound}\Label{saiki}
\subsection{The lower bound}\Label{411}

Let $M_{(1)},..., M_{(n)}$ be a sequence of the POMs in ${\cal H}$,
and 
apply the measurement $M_{(1)}$  to the first sample,
and  the measurement  $M_{(2)}$  to the second sample, and so on.
The choice of $M_(k)$ is dependent on the outcome
$\vec{\omega}_{k-1}=(\omega_{(1)},..,\omega_{(k-1)})$
of $M_{(1)},..., M_{(k-1)}$. 
To reveal the dependency of $M_{(k)}$ on $\vec{\omega}_{k-1}$,
we write $M_(k)[\vec{\omega}_{k-1}]$.

Let us define  the POM $M_n$ in ${\cal H}^n$ 
which takes value in $\Omega^n$ 
by,
\begin{eqnarray*}
  M_n(B)= 
\int_{\vec{\omega}_n\in B}\bigotimes_{k=1}^n 
     M_{(k)}[\vec{\omega}_{k-1}]({\rm d}\omega_{(k)}).
\end{eqnarray*}
Then  the data $\vec{\omega}_n$ is controled by
the probability distribution 
$P^{M_n}_\theta(B)=\tr\rho_\theta^{\otimes n}M_n(B)$.

The estimator is said to be {\it asymptotically unbiased} if
\begin{eqnarray}
(B_n)^i=\left(B_{\theta}\left(\hat{\theta}_n,M\right)\right)^i 
&:= &\int_{\Omega} \left(\hat{\theta}^i_n (\omega) - \theta^i \right)  
{\rm P}^M_{\theta}(\,{\rm d} \omega)\to 0 ~\hbox{ as } n \to \infty,
\Label{k3}\\
(A_n)^i_j= \left(A_{\theta}\left(\hat{\theta},M\right)\right)
&:=& \frac{\partial}{\partial \theta^j}E^i_{\theta}[\hat\theta_n,\,M_n] 
\to \delta^i_j ~\hbox{ as } n \to \infty.\Label{20}
\end{eqnarray}
The MSE consistent estimator satisfies $(\ref{k3})$ always.
Therefore, if apropriate regularity condisions are assumed
so that the defferential, the integral and the trace commute with each other,
then 
$(\ref{20})$ is also satisfied, and the estimator will be 
asymptotically unbiased.  

\begin{thm}\Label{teiri}
If $\{(\hat\theta_n,\, M_n)\}$ is MSE consistent, 
and 
$\vliminf_{n \to \infty} n {\rm V}_{\theta}[(\hat\theta_n,\, M_n)]$ exists,
\begin{eqnarray}
\vliminf_{n \to \infty} n \Tr G {\rm V}_{\theta}[(\hat\theta_n,\,M_n)]
\ge C_{\theta}(G), \Label{siki}
\end{eqnarray}
\end{thm}
\begin{pf}
In the almost same manner as classical estimation theory,
$(\ref{20})$ leads to,
\begin{eqnarray}
n {\rm V}_{\theta}[\hat\theta_n,\,M_n]
\ge n A_n \left( J_{\theta}^{M_n} \right)^{-1} ~^t A_n 
\Label{21}
\end{eqnarray}
Elementary calculation leads to,
\begin{eqnarray}
 \frac{1}{n} J_{\theta}^{M_n}
 = J_{\theta}^{M^n_{\theta}},
\Label{100}
\end{eqnarray}
where $M^n_{\theta} \in {\cal M}$ is a POM in ${\cal H}$ wich id defined by
\begin{eqnarray*}
M^n_{\theta}(\prod_{k=1}^n B_k)=
\int_{\vec{\omega}_n} \sum_{k=1}^n 
  M_{(k)}[\vec{\omega}_{k-1}](B_k)
  {\rm P}_{\theta}^{M_{n}}(\,{\rm d}\vec{\omega}_{n}\,).
\end{eqnarray*}
$(\ref{21})$ and $(\ref{100})$ yield
\begin{eqnarray}
\Tr G n {\rm V}_{\theta}[\hat\theta_n,\,M_n]
\ge \Tr G A_n \left( J_{\theta}^{M^n_{\theta}} \right)^{-1} ~^t A_n 
\ge C_{\theta}(~^t A_n G A_n)  \Label{22}.
\end{eqnarray}
Passing both sides of $(\ref{22})$ to the limit $n\to\infty$,
we have the theorem.
\end{pf}

\subsection{Estimator which achieves the bound}
\Label{kagen}
The estimator defined in the following achives the equality
in the inequality $(\ref{siki})$ if 
the regularity conditions {\rm (B.1-4)} are satisfied.
The proof will be presented later in the subsection \ref{subsec:k2}.
\begin{quote}
First, apply the measurment $M_0$
to $\sqrt{n}$ samples of unknown state $\rho_\theta$, and
calculate 
$\check\theta_n$ which satisfies $(\ref{k1})$.
Second, apply the measurement $M_{\check{\theta}_{n}}$ 
to the remaining $n-\sqrt{n}$ samples,
where  $M_{\theta}$ is defined by
\begin{eqnarray}
\Tr G \left(J_{\theta}^{M_{\theta}}\right)^{-1} 
\leq C_{\theta}(G)+\epsilon',
\Label{k2}
\end{eqnarray}
$(\ref{k2})$ is satisfied. 
Then, 
$\hat\theta_n$ is defined to be $\overline{\theta}_n(\check{\theta}_n)$,
where $\overline{\theta}_n(\theta')$ is defined by,
\begin{eqnarray*}
\overline{\theta}_n(\theta')=
\argmax_{\theta\in\Theta}
\sum_{k=\sqrt{n}+1}^n
\log\frac{{\rm dP}_{\theta}^{M_{\theta'}}}{{\rm d}\nu}(\omega_k).
\end{eqnarray*} 
\end{quote}

\subsection{Regularity conditions}
\Label{subsec:k1}
\begin{itemize}
\item[(B.1)]
There is a POM $M_0$ and 
$\check\theta_n$ which satisfies
\begin{eqnarray}
\lim_{n\rightarrow\infty}
{\rm P}_{\theta}^{ M_n}\{\|\theta-\check{\theta}_n\| \,> \delta \}
=0,\quad \forall \delta>0.
\Label{k1}
\end{eqnarray}
\item[(B.2)] $K:=\sup_{\theta\in\Theta}\|\theta\|$ is finite.
\item[(B.3)]
$\overline{\theta}_n(\theta')$ achieves 
the equality in classical asymptotic Cram{\'e}r-Rao inequality
of the familily $\{{\rm P}_{\theta}^{M_{\theta'}}\,|\,\theta\in\Theta\}$
of probability distributions.
\item[(B.4)]
The higher order term of
mean square error of $\overline{\theta}_n(\theta')$
is  uniformly bounded when $\|\theta'-\theta\|<\delta_1$ 
for some $\delta>0$. In other words,
for any  $\epsilon \,> 0, \theta \in \Theta$,
there exists a positive real number $\delta_1 \,>0$ and  a nutural number
$N$ such that,
\begin{eqnarray}
 \left|
(n-\sqrt{n}) \Tr G {\rm V}_{\theta,n}
- \Tr G \left( J_{\theta}^{M_{\check{\theta}}} \right)^{-1} \right|
\,< \epsilon+\epsilon',\quad
\forall n\ge N,\;\;\forall \check{\theta}\;\; s.t.\;\;
\|\theta-\check{\theta}\| \le \delta_1,
\Label{k4}
\end{eqnarray}
where ${\rm V}_{\theta,n}$ is the conditional mean square error matrix
of $\hat\theta_n$ when $\check\theta_n$ is given. 
\item[(B.5)]
For any $\epsilon \,>0, \theta \in \Theta$,
there exists  $\delta_2 \,> 0, $ such that,
\begin{eqnarray}
\left| \Tr G \left( J_{\theta}^{M_{\check{\theta}}} \right)^{-1}
- C_{\theta}(G) \right|
\,< \epsilon+\epsilon' , \quad 
\forall \theta,\;\forall \check{\theta}\;\; s.t.\;\;
\| \theta- \check{\theta}\| \,< \delta_2.
\Label{k5}
\end{eqnarray}
\end{itemize}

{\rm (B.1)} is satisfied almost always,
and {\rm  (B.2)} is not restrictive .
For $\overline{\theta}_n(\theta')$ is 
the maximum likelihood estimator of the family 
$\{{\rm P}^{M_{\theta'}}_\theta\}$ of probability distributions, 
{\rm (B.3)} is satisfied in usual cases.
The validity of {\rm (B.4)}, however,  is hard to verify.
Therefore, in the future,
this condition needs to be replaced by other conditions. 
Obviously, {\rm (B.5)} reduces to the following {\rm (B.5.1-2)},
both of which are natural.
\begin{itemize}
\item[(B.5.1)]
The map  $\theta \mapsto C_{\theta}(G)$ is continuous.
\item[(B.5.2)]
For any  $\theta'$, the map $\theta\mapsto [J^{M_{\theta'}}_\theta]^{-1}$
is continuous.
\end{itemize}

\subsection{Proof of achivability}
\Label{subsec:k2}
\begin{thm}\Label{te3}
If the model ${\cal M}$ satisfy conditions {\it (B.1-5)} in the following,
then we have,
\begin{eqnarray}
\lim_{n \to \infty} n \Tr G {\rm V}_{\theta}[\hat\theta_n,\,M_n]
= C_{\theta}(G) ,\,\forall \theta \in \Theta,
\Label{48}
\end{eqnarray}
\end{thm}

\begin{pf}
Let us choose $\delta_1,\delta_2$ and $N$ so that $(\ref{k4}-\ref{k5})$
are satisfied, and define  $\delta':= \min(\delta_1,\delta_2)$.
Then, if $n\ge N$, we have,
\begin{eqnarray*}
&&\quad n \Tr G {\rm V}_{\theta}[\hat\theta_n,\,M_n]\\
&&=
n \int
\Tr G {\rm V}_{\theta}[\overline{\theta}_n(\check\theta_n),\,M_n ]
{\rm P}_\theta^{M_n}( \,{\rm d} \omega) \\
&&\le
n \int_{\|\theta-\check{\theta}_n\| \le \delta'}
\Tr G {\rm V}_{\theta}[\overline{\theta}_n(\check\theta_n),\,M_n ]
{\rm P}_\theta^{M_n}( \,{\rm d} \omega)
+ K^2 \Tr G
\int_{\|\theta-\check{\theta}_n\|\,> \delta'} 
{\rm P}_\theta^{ M_n }( \,{\rm d} \omega) \\
&&\le
\frac{n}{n-\sqrt{n}}\int_{\|\theta-\check{\theta}_n\| \le \delta'}
\left(
 \Tr G \left( J_{\theta}^{M_{\check{\theta}_n}} \right)^{-1}
+ \epsilon +\epsilon'\right)
{\rm P}_\theta^{ M_n }( \,{\rm d} \omega) 
+ n K^2 \Tr G 
{\rm P}_\theta^{ M_n }
\{\|\theta-\check{\theta}_n\| \,> \delta \} \\
&&\le
\frac{n}{n-\sqrt{n}}\int_{\|\theta-\check{\theta}_n\| \le \delta}
\left( C_{\theta}(G) + 2\epsilon +\epsilon'\right) 
{\rm P}_\theta^{ M_n }( \,{\rm d} \omega) 
+ n K^2 \Tr G {\rm P}_\theta^{ M_n }
\{\|\theta-\check{\theta}_n\| \,> \delta \} \\
&&\le
\frac{n}{n-\sqrt{n}}(C_{\theta}(G) + 2\epsilon +\epsilon') 
+ n K^2 \Tr G 
{\rm P}_\theta^{ M_n }\{\|\theta-\check{\theta}_n\| \,> \delta \}.
\end{eqnarray*}
{\rm (B.1)} implies that the third term of last end of the equation 
tends to $n$ as $n\to\infty$. Therefore, we have, for every 
$\epsilon'>0$ and for every $\epsilon>0$,
\begin{eqnarray*}
\lim_{n \to \infty} n \Tr G {\rm V}_{\theta}[\hat\theta_n,\,M_n]
\le C_{\theta}(G) + 2\epsilon+\epsilon'.
\end{eqnarray*}
which leads to the theorem.
\end{pf}

\section{Use of quantum correlation}\Label{kage}
In this section, we consider the minimization of 
asymptotic mean square error where $M_n$ runs every POM which satisfies
MSE consistensy. Physically, this means we allow the use of 
interactions between sapmles. 

So far, we considerd POM which takes value in $\Omega$, 
or the totality of all the possible data. 
Instead, in this section, we consider POM with the values in $\R^d$,
for if $M$ is a POM with values in $\Omega$,
$M\circ\hat\theta^{-1}$ is POM with the values in $\R^d$.
MSE consistensy is defined in the same way as the precedent  sections.

Let $C_{\theta}^n(G)$ denote the quasi-quantum Cram\'{e}r-Rao type bound 
of the family $\{ \rho^{\otimes n}_{\theta} |\theta \in \Theta \}$ 
of density operators in ${\cal H}^{\otimes n}$.
Then, the  quantum Cram\'{e}r-Rao type bound $C_{\theta}^Q(G)$ is
defined by,
\begin{eqnarray*}
C_{\theta}^Q(G):=
\vliminf_{n \to \infty} n C_{\theta}^n(G).
\end{eqnarray*}
For $C_{\theta}(G) \ge n C_{\theta}^{n}(G)$
holds true, we have,
\begin{eqnarray*}
C_{\theta}(G)  \ge C_{\theta}^A(G). 
\end{eqnarray*}

\begin{thm}
If the sequence $\{ M^n \}_{n=1}^{\infty}$ is MSE consistent, we have,
\begin{eqnarray}
\vliminf_{n \to \infty} n \tr G {\rm V}_{\theta}\left(M^n \right) 
\ge C^Q_{\theta}(G).
\Label{jen}
\end{eqnarray}
\end{thm}
\begin{pf}
In the  almost same manner as the proof of  theorem \ref{teiri}, we have,
\begin{eqnarray*}
{\rm V}_{\theta}\left[M_n\right]
&\ge& A_{n} \left( J_{\theta}^{M^n} \right)^{-1}  ~^t A_{n}, \\
n \Tr G {\rm V}_{\theta}[M_n]
&\ge& n \tr G A_{n} \left( J_{\theta}^{M_n} \right)^{-1} ~^t A_{n}  \\
& \ge& n C_{\theta}^n\left (~^t A_{n} G A_{n} \right),
\end{eqnarray*}
which approaches $(\ref{jen})$ as $n\to \infty$.
\end{pf}

If  the family
$\{ \rho_{\theta}^{\otimes n} | \theta \in \Theta \}$
of density operators satisfies 
{\rm (B.1-5)}, we have the following theorem.

\begin{thm}
There is a MSE consistent sequence $\{ M_n\}$ of POM 
such that
$\lim_{n \to \infty} n \tr G {\rm V}_{\theta}[M_n]\leq C_{\theta}^Q(G)
+ \epsilon$ 
is satisfied for every $\epsilon>0$ and for every $\theta\in\Theta$.
\Label{prodtassei}
\end{thm}
\begin{pf}

Let us devide $n$ samples into $n_2$ groups each of which 
is consist of $n_1$ samples,
and let $M_{(1)}^{n_1},\ldots, M_{(n_2)}^{n_1}$ be
a sequence of POMs in ${\cal H}^{\otimes n_1}$.
Apply the measurement $M_{(1)}^{n_1}$ to the first group 
$\rho_\theta^{\otimes n_1}$ of samples, 
and apply $M_{(2)}^{n_1}$ to the second samples, and so on.
The choice of $M_{(k)}^{n_1}$ is dependent on the outcome of 
the measurements  $M_{(1)}^{n_1},\ldots, M_{(k-1)}^{n_1} $.
With $n_1$ fixed, let us approach $n_2$ to $\infty$. Then,  
theorem \ref{te3}implies the existence of
 a MSE consistent sequence $\{M_n\}$
of POM  which satisfies 
\begin{eqnarray}
\lim_{n \to \infty} n \tr G {\rm V}_{\theta}[M_n]= 
\lim_{n_2 \to \infty} n_1 n_2 \tr G {\rm V}_{\theta}[M_n]=
n_1 C_{\theta}^{n_1}(G).
\end{eqnarray}
For any epsilon, if $n_1$ is sufficiently large,  
$\lim_{n \to \infty} n \tr G {\rm V}_{\theta}
\left(M^n \right)\leq  C_{\theta}^Q(G)
+ \epsilon$ is satisfied, and we have the theorem.
\end{pf}

\end{document}